\documentclass[twocolumn,showpacs,amsmath,amssymb]{revtex4}
        
\usepackage{graphicx}
\usepackage{dcolumn}
\usepackage{bm}
\usepackage{bbm}
\usepackage{amsfonts}

\def\Z{{\mathbb{Z}}}

\def\>{\rangle}
\def\<{\langle}
\def\ket#1{|#1\>}
\def\bra#1{\<#1|}

\def\ave#1{\left\< #1\right\>}

\def\tr{{\,{\rm tr}}}

\def\ii{{\rm i}}
\def\dd{{\rm d}}

\def\LL{{\cal\hat L}}

\begin{document}

\title{Negative differential conductivity in Heisenberg XXZ chain far from 
equilibrium\footnote{This note has been written up in February 2004 when the author was
visiting NUS, Singapore. Even though it has not been published, I believe the reported result
is still quite intriguing and awaiting physical explanation.}}
\author{Toma\v z Prosen}
\affiliation{Physics Department, Faculty of Mathematics and Physics, 
University of Ljubljana, Ljubljana, Slovenia}

\begin{abstract}
Negative differential conductivity is reported for the far from equilibrium 
quantum spin transport in the insulating regime ($J_{\rm x} < J_{\rm z}$) 
of finite Heisenberg XXZ spin $1/2$ chains. The phenomenon is reproduced 
analytically for small chains of $N=4$ spins and further analyzed numerically,
for up to $N=16$, using an efficient pure-state simulation with
stochastic spin baths.
\end{abstract}

\maketitle

Quantum transport properties of low dimensional materials are currently an
object of intensive theoretical (see Ref.~\cite{1dqtr} for a recent
review) and experimental\cite{experiments} research. 
Particularly interesting is the case of quasi-1d strongly
interacting particle systems, where an important link between conductivity 
and integrability has been recently established \cite{zotos}. However, most 
of theoretical studies focused on the close to equilibrium situation by 
using the {\em linear response} formalism,
and almost nothing is known about the physics of such systems
{\em far from equilibrium} (FFE).
In order to drive a small interacting quantum system FFE one has to couple 
it strongly to some macroscopic reservoirs. Theoretical description of this 
situation usually goes via master equation
for the density matrix where the non-unitary (dissipative) term depends
on the coupling of the model to the reservoirs. Numerical simulations of such
situations in non-trivial models have only recently became computationally
feasible \cite{saito}.

In this paper we propose conceptually simple and perhaps experimentally
realizable form of coupling of a small 1d interacting
quantum system to a pair of macroscopic baths of spins (or spinless fermions, 
or any other quantum two level systems -- qubits). Our setting can also 
be viewed as a simple model
of the qubit transport which may be of relevance in quantum information.
In addition it allows for a very efficient (stochastic) numerical 
simulation of the non-equilibrium steady state (NESS) in terms of a pure 
state which only after averaging over stochastic bath interactions 
statistically converges to the proper density-matrix of NESS.
We believe that, for a generic, non-pathological quantum interacting system, 
the bulk properties of NESS in the thermodynamic limit should not depend
on the model of the baths. And now we come to the main point.
We apply our model to simulate FFE spin transport in the well known
Heisenberg XXZ spin $1/2$ chain. While in the regime, known
as {\em ideally conducting} \cite{zotos,xxz}, we find expected results, 
namely that the spin current increases monotonically (and almost linearly) 
with the increasing driving field, we find a very different
result in the other regime, which is for zero magnetization and
close to equilibrium known as an {\em ideal insulator}. Namely there the
spin (or qubit) current appears to have a clear maximum as a function of the 
driving field, hence for sufficiently strong field we find that the current
decreases upon the increase of driving, and practically vanishes
for a maximal field. This result can be reproduced analytically
for a small chain of $N=4$ spins, while numerical simulations up to 
$N=16$ indicate that it becomes even shaper by increasing $N$.
Negative differential conductivity has been theoretically predicted
and observed before, mainly in semiconductors (see e.g. Ref.~\cite{ndc}),
as a consequence of various dynamical current instabilities.
However, we believe that our result provides a new paradigm for such a
behavior in strongly interacting quantum many-body systems
with possible applications in the transport of quantum information.

We begin by describing our general setting of system-bath coupling.
Let our systems consist of $N$ spins $1/2$, or qubits, described by Pauli 
variables $\vec{\sigma}_n$, $n=1,\ldots,N$. The first and the last spin, 
$\vec{\sigma_1}$ and $\vec{\sigma}_N$, shall be coupled to the baths, 
so let us decompose the total $2^N$ dimensional Hilbert space to $2^{N-2}$ 
dimensional Hilbert space of {\em interior} chain and $4$ dimensional 
space of {\em border} qubits, ${\cal H} = {\cal H}_{\rm in} \otimes 
{\cal H}_{\rm bo}$. Let the left and the right bath be characterized by
some chemical potentials $\mu_{\rm L},\mu_{\rm R}\in [0,1]$, and
let $U(t)$ denote $2^N$ dimensional unitary evolution matrix of an
autonomous spin chain for duration $t$. Usually we can write 
$U(t) = \exp(-\ii t H)$ in terms of some Hamiltonian matrix $H$, but working 
with a unitary map $U(t)$ may be more general, e.g. representing also
periodically
kicked (or driven) systems (such as in Ref.~\cite{ktv}). 
Concerning the evolution
$U(t)$ we shall only assume that it conserves the total magnetization
$M=\sum_{n=1}^N \sigma^{\rm z}_n$, {\em i.e.} the qubit number $(M+N)/2$ if 
spin {\em up} represents a qubit state, namely $[U(t),M]=0$. 
Let the canonical (qubit) basis $\ket{c}$ of ${\cal H}$, {\em i.e.} eigenbasis
of $\sigma^{\rm z}_n$, be labelled by $N$-digit binary decompositions 
$c=\sum_{n=1}^N c_n 2^{n-1}$, $c_n\in\{0,1\}$, 
where $2c_n - 1$ is an eigenvalue of $\sigma^{\rm z}_n$.
Any $\ket{c}$ can be written as a direct product 
$\ket{c} = \ket{a}_{\rm in}\otimes \ket{b}_{\rm bo}$ where
$a =\sum_{n=2}^{N-1} c_n 2^{n-2}$ and $b = c_1 + 2 c_N$.
Conversely, we define $\gamma(a,b):=c$.

We shall now describe the system's evolution when it is coupled to the baths.
Consider an ordered sequence of times $t_k,k\in\Z$, $t_k < t_l$ for $k < l$,
such that $\lim_{k\to\pm\infty}t_k = \pm\infty$. 
Between two subsequent times, say $t_{k-1}$ and $t_k$, the evolution
is described by a unitary propagator $U(t_k-t_{k-1})$.
Then, at time $t_k$ we perform a measurement of the z-component of the
first and the last spin,
$\sigma^{\rm z}_{1},\sigma^{\rm z}_N$. If $\ket{\psi} = \sum_c \psi_c \ket{c}$
is a pure state drawn from a statistical ensemble describing a state just
before the measurement then, after the measurement, the state collapses
with probability $p_b=\sum_a|\psi_{\gamma(a,b)}|^2$ to one
of the four pure states $\ket{\psi_b}=\left(\sum_a \psi_{\gamma(a,b)}
\ket{a}_{\rm in}\right) \otimes \ket{b}_{\rm bo}$. After the measurement,
since the states of the two border spins are known, we can adjust their
expectation value to the ones of the reservoirs. This is simply achieved 
by applying, conditionally, spin-flips, i.e. one-qubit gates 
$\sigma^{\rm x}_{1}$ (or $\sigma^{\rm x}_N$), such that after the flips the probability that the 
left (right) spin points up is exactly $\mu_{\rm L}$ ($\mu_{\rm R}$).
In a practical Monte-Carlo simulation, this means that we generate two
uniform random numbers $\zeta_{\rm L},\zeta_{\rm R}\in [0,1]$.
If $\zeta_{\rm L,R} < \mu_{\rm L,R}$ we require that $c_{1,N}=1$ so we
apply $\sigma^{\rm x}_{1,N}$ only if $b_{1,2}=0$, where $b=b_1 + 2 b_2$.
On the contrary, if $\zeta_{\rm L,R} > \mu_{\rm L,R}$ we require that 
$c_{1,N}=0$ so we apply $\sigma^{\rm x}_{1,N}$ only if $b_{1,2}=1$. Since 
conditional spin flips only affect the states of the border spins
$\ket{b}_{\rm bo}\to\ket{b'}_{\rm bo}$, the state
at time $t_k+0$ is still a direct product 
$\ket{\psi'}=\left(\sum_a \psi_{\gamma(a,b)}\ket{a}_{\rm in}\right)
\otimes\ket{b'}_{\rm bo}$. Then we continue with autonomous evolution
$U(t_{k+1}-t_{k})$ to the next instant $t_{k+1}$ 
and repeat the probabilistic procedure of
measurement and conditional spin flips.
Provided that the distribution of time lags 
$\tau_k:=t_{k+1}-t_{k}$ has stationary 
statistical properties we conjecture that a well defined NESS
is approached as $t_k\to\infty$. This is evident in case 
$\tau_k=\tau={\rm const}$ 
which shall mostly be studied below. During such a simulation, an average
spin (or qubit) current $j_{\rm L}$ is calculated by looking at the 
left bath and summing up all down-up spin flips minus all up-down flips and 
dividing by the time of simulation. Due to conservation of $M$ this 
current should be after long time precisely equal to the analogous quantity 
$j_{\rm R}$ at the right bath.

The procedure described above is very suitable for efficient numerical method
which is by far superior to any density matrix simulations since one has
to deal only with $2^N$ dimensional vector $\psi_c$ whereas in solving
master equations we have to treat the full $2^N\times 2^N$ density matrices.
However, for analytical treatment, the above procedure can be
written by means of probabilistic ensembles and density matrices, {\em i.e.}
in terms of an effective master equation. Statistical state averaged over
an ensemble of bath interactions $\ave{\ket{\psi'}\bra{\psi'}}_{\rm bath}$
at time $t_k+0$ can always be written as a direct product 
$\rho_k\otimes \omega$ where 
$\omega = \sum_{b=0}^3 \omega_b \ket{b}_{\rm bo}\bra{b}_{\rm bo}$,
$\omega_b = \frac{1}{4}(1+(-1)^{b_1}\mu_{\rm L})(1+(-1)^{b_2}\mu_{\rm R})$ is the
density matrix of the border spins controlled by the baths.
Writing the dynamical equation for the interior part $\rho_k$ is straightforward.
If $P_b : {\cal H}_{\rm in} \to {\cal H}$ denotes the lift
$P_b\ket{\psi}_{\rm in} := \ket{\psi}_{\rm in}\otimes \ket{b}_{\rm bo}$ and
$P_b^\dagger : {\cal H} \to {\cal H}_{\rm in}$ the corresponding projection, then
\begin{equation}
\rho_{k+1} = \sum_b P^\dagger_b U(\tau_k) 
(\rho_k \otimes \omega) U^\dagger(\tau_k) P_b.
\end{equation}
Assuming the simple case $\tau_k={\rm const}$, and writing $U=U(\tau)$, 
NESS $\rho_\infty$ 
is an eigenvalue $1$ eigenvector of a completely positive master operator
\begin{equation}
\LL\rho := \sum_b P^\dagger_b U (\rho\otimes\omega) U^\dagger P_b,
\end{equation}
namely
\begin{equation}
\LL\rho_{\infty} = \rho_{\infty},
\label{eq:fixpoint}
\end{equation}
and is unique provided $1$ is a non-degenerate eigenvalue.
The next largest eigenvalue of $\LL$ would determine the rate of relaxation
of an arbitrary initial state to NESS $\rho_\infty$.
To determine the spin current 
$\ave{j}_\infty$ in the steady state we have to compute the
probabilities of measurements
\begin{equation}
p_b = {\rm tr\,} P^\dagger_b U (\rho_\infty \otimes \omega) U^\dagger P_b
\end{equation}
and then, considering either left or right bath,
\begin{equation}
\ave{j}_\infty = \frac{p_0 + p_2 - \omega_0 - \omega_2}{\tau} = 
\frac{\omega_0 + \omega_1 - p_0 - p_1}{\tau}.
\end{equation}

In the following we concentrate on analytical calculations and numerical
simulations in a specific important model, namely the anisotropic Heisenberg XXZ
spin chain with the Hamiltonian
\begin{equation}
H = \sum_{n=1}^{N-1}\left[J_{\rm x}(\sigma^{\rm x}_n \sigma^{\rm x}_{n+1}
+ \sigma^{\rm y}_n \sigma^{\rm y}_{n+1}) + J_{\rm z}\sigma^{\rm z}_n \sigma^{\rm z}_{n+1}
\right].
\end{equation}
The autonomous model is completely solvable (either with open or
periodic boundary conditions) in terms of Bethe Ansatz (see e.g.\cite{korepin}),
and its thermodynamic ($N\to\infty$) transport properties are determined \cite{xxz} by 
$\Delta=J_{\rm z}/J_{\rm x}$, and $s = M/N$. The model is {\em ideal conductor}, i.e.
d.c. conductivity diverges for $s \neq 0$, whereas for $s = 0$, also corresponding
to thermodynamical grand canonical state, the model shows a transition from
ideal conductor for $|\Delta| < 1$ to {\em ideal insulator} for $|\Delta| > 1$.
All the classical results about this model are based on linear response calculation
and thus refer to the close to equilibrium situation. In this paper we analyze
its FFE properties with the model of spin baths described above.

First we perform numerical Monte-Carlo simulations of quantum NESS 
exactly as described above.
We choose a symmetric driving, $\mu_{\rm L,R} = \frac{1}{2}(1\mp \mu)$ so
$\mu=\mu_{\rm R}-\mu_{\rm L} \in [0,1]$ is a parameter controlling the `field strength'.
$\mu=1$ corresponds to the maximum driving, where after the bath-interaction 
left/right-most spin is always pointing down/up.
We always started from an initial random state, where coefficients $\psi_c$ were chosen
as complex random Gaussian numbers of equal variance. Then we performed $10^6$ steps
of integration, while first $10^4$ steps were omitted to ensure convergence to
steady state. Time averages of all the quantities have been carefully checked for
convergence. Three different choices for the distribution of time-lags $\tau$ have
been chosen: (A) constant time-lag $\tau_k=\tau=1$, (B) uniform distribution
of time-lags $\dd {\cal P}/d\tau=1/2$ in the interval $[0,2]$, 
(C) exponential distribution of time-lags $\dd {\cal P}/d\tau=\exp(-\tau)$
meaning that instants $t_k$ are independent Poissonian events.

\begin{figure}
\vspace{-8.5mm}
\centerline{\includegraphics[width=3.4in]{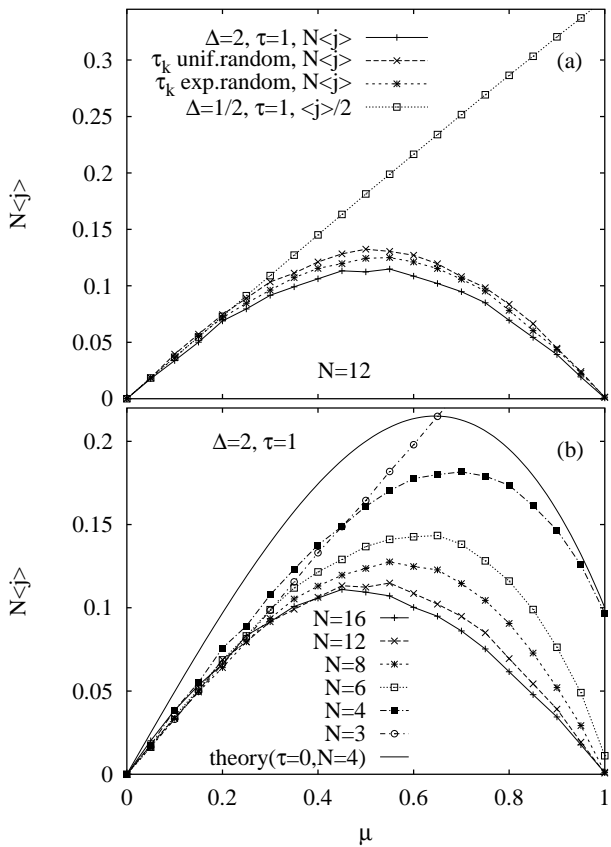}}
\caption{Current $N\ave{j}$ as determined from numerical simulations is plotted versus 
field strength $\mu$.
In (a) we compare, for $N=12$, ideally conducting ($\Delta=1/2$) and
insulating ($\Delta=2$) cases.
In the first case we only show simulation with uniform time-lag $\tau=1$ while
in the second, we compare results for three different distributions of $\tau_k$ (see text).
In (b) we concentrate on the anomalous case ($\Delta=2$) and compare different sizes $N$
(all for uniform $\tau_k=\tau=1$). Full curve is the theoretical formula 
(\ref{eq:j4}) for $N=4,\tau\to 0$, divided by $\tau$. 
}
\label{fig:jm}
\end{figure}

In Fig.~\ref{fig:jm}a we show steady state current $\ave{j}=j_{\rm L}=j_{\rm R}$ as
a function of the field strength $\mu$ for $N=12$. For $\Delta=1/2$ 
($J_{\rm x}=1,J_{\rm z}=1/2)$), we find almost perfectly linear 
behavior $\ave{j}\propto\mu$ 
meaning that linear response
results can be extended to arbitrary field strength. Repeating the simulation
for other values of $N=8,16$ and fixed $\mu=0.3$, we find also that the current is
almost independent of $N$, namely
$\ave{j}|_{N=8}=0.2198$, $\ave{j}|_{N=12}=0.2182$, $\ave{j}|_{N=16}=0.2187$, 
which is consistent with ideal conductivity.

The we turn into the insulating regime and put $\Delta = 2$ while keeping 
$J_{\rm z}=1/2$. Here we find a very strange behavior of $\ave{j}$ as a function
of $\mu$ as seen in Fig.~\ref{fig:jm}a. The current has a maximum around $\mu\approx 1/2$ and
then for further increasing $\mu$ it decreases, so that it essentially vanishes at 
$\mu=1$. This seems extremely surprising phenomenon for which we have no intuitive 
explanation. In order to exclude possible interference effects due to
periodic bath-interactions we repeat the same simulation for randomized time lags,
(B) and (C), and we find qualitatively exactly the same behavior (fig.~\ref{fig:jm}a).

In Fig.~\ref{fig:jm}b we plot the current-field characteristics for different
system sizes up to $N=16$. We find that for small $\mu$, current is in fact
proportional to the field gradient $\mu/N$, like for a normal conductor obeying
some diffusion law. However, this cannot hold in the thermodynamics limit, since
we know that for $N=\infty$ the system should behave as an insulator. 
Indeed, we observe that the maxima of the curves $\ave{j}(\mu)$ decrease towards
smaller $\mu$ as we increase $N$. 
Yet, the behavior of current-field characteristic for a finite value of $N$ is
highly intriguing and requires further analysis and explanations.

\begin{figure}
\centerline{\includegraphics[width=3.4in]{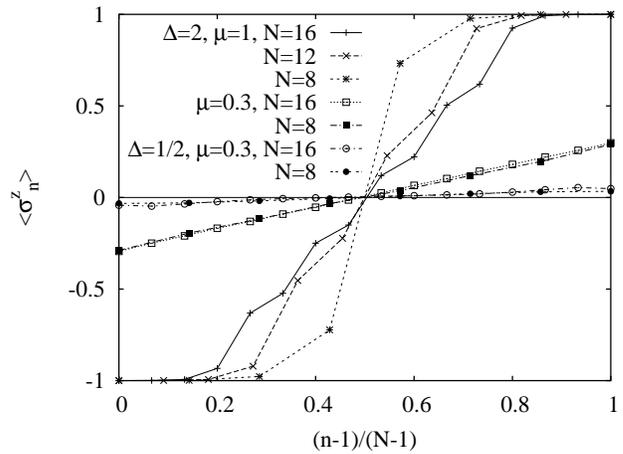}}
\caption{We show spin density profiles $\ave{\sigma^{\rm z}_n}$ versus the
scaled coordinate $(n-1)/(N-1)$, as determined from numerical simulations,
for ideally conducting ($\Delta=1/2$) and insulating ($\Delta=2$) cases, 
at small field $\mu=0.3$, and for the insulating case
at large field $\mu=1$. In all three cases, results for several different sizes $N$
are compared.
}
\label{fig:sn}
\end{figure}

During the simulations we have also computed the spin-density profile, namely the
time-averages of expectation value $\ave{\sigma^{\rm z}_n}$ as a function of the
lattice index. 
For ideal conductor we expect that no density-gradient can be build and
indeed we find very flat spin-density profile for the case $\Delta=1/2$
shown in Fig.~\ref{fig:sn}. Further we show in the same figure two density profiles
for the insulating case $\Delta=2$, one in the regime
left to the maximum of current-field characteristic, namely for $\mu=0.3$, and one
for the extreme field $\mu=1$ where the current is practically zero.
We observe that for $\mu=0.3$ the density-profile is very nicely scaling with 
$(n-1)/(N-1)$ indicating that the system can support a linear density-gradient like
one observes in normal conductor. Indeed in this regime the system is indistinguishable
from a normal conductor. Still, for strong filed $\mu=1$, different density-profile
is observed with spin-density changing much slower near the baths than in the bulk.
In this regime it seems difficult to define a thermodynamic density gradient.

\begin{figure}
\centerline{\includegraphics[width=3.4in]{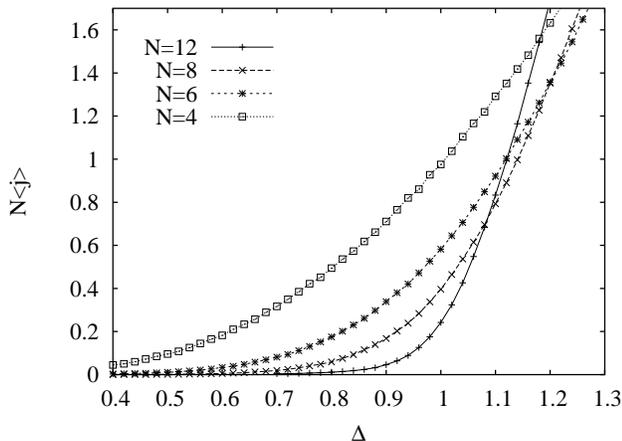}}
\caption{Spin current at maximum field, $N\ave{j}|_{\mu=1}$, is plotted as
a function of $\Delta=J_{\rm z}/J_{\rm x}$, at fixed $J_{\rm z}=1/2$ and uniform
$\tau_k=\tau=1$, for four different sizes $N$.
}
\label{fig:jj}
\end{figure}

Finally, we suggest to use the current at maximum field $\theta=N\ave{j}_{\mu=1}$ as
an order parameter signaling a transition form a {\em normal} behavior $\theta > 0$
for $|\Delta| < 1$ to {\em anomalous} behavior $\theta=0$ found for $|\Delta|> 1$. 
In Fig.~\ref{fig:jj} we show the dependence $\theta(\Delta)$ for fixed $J_{\rm z}=1/2$,
and for different values of $N$, and indeed we find that for increasing $N$, the
transition becomes increasingly abrupt at the critical parameter $\Delta=1$.

We note that, for other values of parameters $J_{\rm x}$ and $J_{\rm z}$,
$\ave{\tau_k}$, qualitatively similar numerical results were obtained, 
essentially depending only on the ratio $\Delta$ apart from trivial scaling factors.

Does this phenomenon allow for a simple analytical description, 
at least for a small system? From Fig.~\ref{fig:jm}b we learn that the
phenomenon exist already for $N=4$ (and still not for $N=3$), for which we might hope to 
solve analytically the fixed point equation (\ref{eq:fixpoint}). Indeed it turns
out that $N=4$ is the maximal dimension which allows for explicit solution of
eq. (\ref{eq:fixpoint}), however only asymptotically for small $\tau$. 
This is due to the
fact that the propagator $U$ cannot be written in a closed form so  
$\exp(-\ii \tau H)$ has to be expanded to second order in $\tau$.
Then it turns that the $16\times 16$ matrix of $\LL$ indeed has a unique 
right eigenvector $\rho_\infty$ of maximum eigenvalue $1$. It is interesting to 
note that within first order in $\tau$ the matrix of $\rho_\infty$ is {\rm real} 
and the lowest order contribution to the steady state current is only in $O(\tau^2)$.
Lengthy but straightforward calculations have been performed by means of Mathematica.
The final solution for the current reads
\begin{equation}
\ave{j}_\infty = \frac{2\tau J_{\rm x}^2(\mu_{\rm R}-\mu_{\rm L})(1 + 
2(\mu_{\rm L}+\mu_{\rm R}-\mu_{\rm L}^2-\mu_{\rm R}^2)\Delta^2)}{1 + 
(\mu_{\rm L}+\mu_{\rm R})(2-\mu_{\rm L}-\mu_{\rm R})\Delta^2}
\label{eq:j4}
\end{equation}
with the correction $O(\tau^3)$.
Note that vanishing of $\ave{j}_\infty$ as $\tau\to 0$ is just a
manifestation of quantum Zeno effect when we approach the limit of 
{\em continuous measurement}.
For the symmetric case, $\mu_{\rm L}+\mu_{\rm R}=1,\mu_{\rm R}-\mu_{\rm L}=\mu$, 
this formula is plotted in Fig.~\ref{fig:jm}b, with a maximum at 
$\mu = \sqrt{(1+\Delta^2)/3}$, and is not far
from numerical simulation for $\tau=1$.
While the spin density at the border is
fixed by the baths, its interior gradient can be directly computed 
$\ave{\nabla\sigma^{\rm z}}_\infty
={\rm \tr}[(\sigma^{\rm z}_3-\sigma^{\rm z}_2)\rho_{\infty}]$,
\begin{equation}
\ave{\nabla\sigma^{\rm z}}_\infty = 
\frac{2 (\mu_{\rm R}-\mu_{\rm L})^3 \Delta^2}{1 + 
(\mu_{\rm L}+\mu_{\rm R})(2-\mu_{\rm L}-\mu_{\rm R})\Delta^2}
\end{equation}
and is always monotonic, though cubic, function of the
field strength $\mu_{\rm R}-\mu_{\rm L}$.
This means that not only the external field strength is increasing, but
also the interior magnetization gradient is increasing while the
current is decreasing, for large $\mu$.

In conclusion, we have studied qubit (or spin) quantum transport
in finite interacting systems far from equilibrium, for which a special
efficient model of macroscopic baths has been designed. 
The interpretation of this model is very simple: at periodic, or
randomly chosen, instants of time we ``look at'' (measure) 
the border 2-level states of the system, whether they are occupied
or not. Then we conditionally flip this border state (i.e. exchange
the qubit with the bath) so that after this
process the density of qubits (or spin magnetization) at the
borders is prescribed. 
Applying this model to study FFE transport in finite
Heisenberg XXZ chain, we find negative differential conductivity for 
$\Delta > 1$ provided the spatially varying magnetization of NESS 
crosses the insulating point $s=0$. In this regime, 
the optimal quantum strategy to transport 
qubits from left to right is not simply filling them from one end and taking
out from the other, but it is better to sometimes insert 0-qubit state from the
left and not taking 1-qubit state from the right.

The author acknowledges discussions with G. Casati, P. Prelov\v sek and 
A. Ram\v sak, support by the grant P1-0044 of MZV\v S, Slovenia, and
hospitality of National University of Singapore where this work has been
completed.

\end{document}